\documentclass[aps,prc,showpacs,twocolumn,preprintnumbers,floatfix,
showkeys,tightenlines]{revtex4}
\usepackage{amsmath,amssymb,amsfonts}
\usepackage{graphicx,psfig}
\usepackage{color}
\allowdisplaybreaks

\newcommand{\be}{\begin{equation}}
\newcommand{\ee}{\end{equation}}
\newcommand{\bea}{\begin{eqnarray}}
\newcommand{\eea}{\end{eqnarray}}
\newcommand{\bm}{\bibitem}

\newcommand{\al}{\alpha}

\newcommand{\ep}{\epsilon}
\newcommand{\de}{\delta}
\newcommand{\De}{\Delta}
\newcommand{\om}{\omega}

\newcommand{\lm}{\lambda}

\newcommand{\sg}{\sigma}

\newcommand{\ze}{\zeta}

\newcommand{\cz}{{\cal Z}}

\begin{document}

\setcounter{page}{1}

\title{$S$-matrix approach to equation of state of nuclear matter}

\author{S. \surname{Mallik}}
\email{mallik@theory.saha.ernet.in}
\author{J.N. \surname{De}}
\email{jn.de@saha.ac.in}
\author{S.K. \surname{Samaddar}}
\email{santosh.samaddar@saha.ac.in}
\affiliation{Theory Divsion, Saha Institute of Nuclear Physics, 1/AF 
 Bidhannagar, Kolkata 700064.}  
\author{Sourav \surname{Sarkar}}
\email{sourav@veccal.ernet.in}
\affiliation{Variable Energy Cyclotron Centre, 1/AF, Bidhannagar, 
Kolkata, 700064,India}

\date{\today}

\begin{abstract} 
We calculate the equation of state of nuclear matter based on the general analysis 
of the grand canonical partition function in the $S$-matrix framework. In addition 
to the low mass stable particles and their two-body scattering channels considered 
earlier, the calculation includes systematically all the higher mass particles and 
their exited states as well as the scattering channels formed by any number of these 
species. We estimate the latter contribution by resonances in all the channels. 
The resulting model-independent virial series for pressure gets substantial 
contribution from the heavy particles and the channels containing them. The series 
converges for larger values of baryon density than found earlier. 
\end{abstract}


\maketitle

The equation of state of nuclear matter in the hadronic phase is of central interest
in the study of supernovae and neutron star matter in astrophysical context and of 
heavy ion collisions in the laboratory. There are microscopic frameworks to deal 
with the nuclear strong interaction involved here, such as the variational 
\cite{Pan,Akmal} and the Brueckner-Goldstone \cite{Haar,Jensen,Baldo} methods, 
using $N\!N$ potentials based on experimental scattering data. There are also  
empirical expressions \cite{Lattimer,De} for the Helmholtz free energy 
using a model for nuclei. However, these approaches are model-dependent, involving 
some theoretical uncertainties.

A more direct and simple approach based completely on observables was suggested long 
ago by Beth and Uhlenbeck \cite{Beth,Huang}, who expressed the second  coefficient 
in the virial expansion for pressure of a gas in terms of two-body scattering 
phase-shifts of its constituents. But it does not shed much light on the virial 
series as a whole, so that a straightforward calculation up to second order may not 
be trusted, except for very dilute systems.

A definitive step in this direction was taken by Dashen, Ma and Bernstein
\cite{Dashen1}, who gave the complete $S$-matrix formulation of statistical 
mechanics. Their work is the logical completion, though in a highly non-trivial way,
of the calculation of the second virial coefficient cited above. 
Starting from a few 'elementary' species in the definition of the  
grand partition function, its calculation ends up with a multi-species system, 
with all the bound states treated on the same footing as the original 
'elementary' species. The complete formula involves two separate pieces. The first 
piece is a sum of ideal gas terms, one for each of these stable species. The 
second piece originates from all possible scatterings of any number of particles 
from these species; it is in the form of a virial sum over
all the scattering channels of terms involving their $S$-matrix elements.
The convergence of the virial series is determined by the fugacities and 
the binding energies of these species.  

An attempt to include a bound state heavier than the two-nucleon system (deuteron) 
in the Beth-Uhlenbeck scheme has been made by Horowitz and Schwenk \cite{Horowitz}. 
They incorporate the alpha particle explicitly in the definition of the grand 
partition function itself along with the proton and the neutron 
and expand it to second order in their fugacities to get both the ideal gas and the 
scattering terms from these particles to that order. As expected, the large 
binding energy of the alpha particle enhances its formation in the system lowering 
the pressure. 
  
In the present work we consider the complete expression for the grand partition 
function \cite{Dashen1}, as applied to nuclear matter, including, in particular, 
all the massive species and the scattering channels formed by them. The only major 
approximation is that of resonance domination of $S$-matrix elements for scattering 
channels with massive particles. As shown by Dashen and Rajaraman 
\cite{Dashen2,Dashen3}, it gives rise to additional ideal gas terms from 
these resonances. 

We consider nuclear matter, consisting of protons and neutrons as two independent 
species of 'elementary' particles, interacting strongly in the limit of isospin 
symmetry. The object of study is the grand canonical partition function for this 
system,
\be
\cz =Tr\, e^{-\beta (H-\mu_p\hat{N}_p -\mu_n\hat{N}_n)}\,,
\ee
where $H$ is the total Hamiltonian, $\hat{N}_{p,n}$ the nucleon number operators, 
$\beta$ the inverse of temperature $T$ of heat bath and $\mu_{p,n}$ the 
nucleon chemical potentials. The trace is taken over any complete set of states 
of all possible numbers of nucleons. Denoting the fugacities by 
$\ze_p=e^{\beta\mu_p},~~ \ze_n=e^{\beta \mu_n}$, the full trace can be decomposed as
\be
\cz = \sum_{Z,N=0}^{\infty} \ze_p^Z \ze_n^N Tr_{Z,N}\, e^{-\beta H}\,,
\ee
where $Tr_{Z,N}$ is now taken over states of $Z$ protons and $N$ neutrons.
For small $\ze_p$ and $\ze_n$, $\ln Z$ can be expanded in a virial double series,
\be
\ln \cz = \sum_{Z,N=1}^{\infty} A_{Z,N}\ze_p^Z \ze_n^N \,.
\ee   
Our task is to calculate the virial coefficients $A_{Z,N}$.

We work in natural units, $\hbar =c=1$.
For later reference, we note here the virial expansion for the ideal quantum gas of 
bosons and of fermions, 
\bea
\ln {\cz}^{(0)} &=& \mp gV\int\!\frac{d^3p}{(2\pi)^3}\,\ln (1 \mp \ze e^{-\beta p^2/2m})
\nonumber\\
&=& \frac{gV}{\lm^3(m)}\left(\ze \pm\frac{\ze^2}{2^{5/2}} +\frac{\ze^3}{3^{5/2}}
        +\cdots\right)\,.
\eea
Here and below the superscript $(0)$ will indicate ideal quantum gas. The upper and
lower signs correspond respectively to bosons and fermions. The degeneracy of the 
single particle states is denoted by $g$ \cite{Comment1}, $V$ is the volume of the 
system and $\lm (m)=\sqrt{2\pi/(mT)}$, the thermal wavelength of a particle of 
mass $m$. The integral representation above is valid for all values of $\ze$, but 
its logarithmic branch point at $\ze=\pm 1$ makes the series converge only for 
$|\ze| < 1$.

Though Coulomb interaction is assumed absent in nuclear matter, we shall include 
its effect on the binding energies of nuclei needed below. Our calculation can 
then be readily applied to a physical system, such as neutron star matter, 
by including the effect of added electrons necessary to make it electrically 
neutral.  

We now follow Dashen {\it et al} \cite{Dashen1} to write the grand partition 
function as the sum of two types of terms,
\be
\ln \cz =\ln {\cz}_{part}^{(0)} + \ln {\cz}_{scat},
\ee
corresponding to contributions from stable, single {\it particle} states and 
(multiparticle) {\it scattering} states respectively.

Let us first concentrate on the particle piece, $\ln Z^{(0)}_{part}$. If $Z$ protons 
and $N$ neutrons form a bound state (nucleus) of mass number $A=Z+N$, it has mass 
$Am$ and energy,
\be
\ep_{Z,N}=\frac{p^2}{2Am} -B_{Z,N}\,, 
\ee
in the ground state, where $\vec{p}$ is its momentum and $B_{Z,N}$ the binding 
energy. From the condition of chemical equilibrium among different species, its 
chemical potential is $\mu_{Z,N}=Z\mu_p +N\mu_n$. Further, all these nuclei, 
particularly the heavy ones, have a large number of dense excited states 
above their ground states, which are stable in the absence of electromagnetic
interaction. (Actually their radiation widths are very small, of the order of eV.) 
Accordingly we split the particle piece in Eq.(5) into two,
\be 
\ln {\cz}_{part}^{(0)}=\ln {\cz}_{gr}^{(0)} +\ln {\cz}_{ex}^{(0)} ,
\ee 
denoting respectively the contribution of the {\it ground} and the {\it excited} 
states.

The first term in Eq.(7) is a sum of ideal gas terms, one for each of the 
ground states of all the nuclei, 
\be
\ln {\cz}_{gr}^{(0)}=\mp V\sum_{Z,N}g \int\!\frac{d^3p}{(2\pi)^3}\, \ln \left( 1\mp
e^{-\beta(p^2/2Am -B_{Z,N} -\mu_{Z,N})} \right) \\
\ee
where $Z$ and $N$ count the number of protons and neutrons in these species.      
As in Eq.(4) the $\mp$ sign correspond to nuclei with $A$ even and odd, obeying 
Bose and Fermi statistics respectively. The sum, of course, includes the original 
elementary particles, namely the proton and the neutron with their $B_{Z,N}=0$.

The second term in Eq.(7) gives the ideal gas terms for each of the elements in
the set of excited states of the nuclei. Here the individual gas terms 
are actually suppressed by an additional Boltzmann factor, $e^{-\beta E}$, $E$ being 
the excitation energy above the ground state. But the density of excited 
states is rather high in heavy nuclei with $A > 8$, say and can, in fact, more than 
compensate for this suppression. We write the contribution of the excited states
of a single nucleus as an integral over $E$ of an ideal gas term multiplied by their 
level density $\om(A,E)$, which we take as \cite{Bohr, Feshbach},
\be
\om (A,E)=\frac{\sqrt{\pi}}{12a^{1/4}}\frac{e^{2\sqrt{aE}}}{E^{5/4}}\,,
\ee
obtained from the Fermi gas model of non-interacting nucleons within the nucleus.
We adopt the empirical value for the parameter, $ a \cong A/8~~(MeV)^{-1}$.
We thus get 
\bea 
&&\ln {\cz}_{ex}^{(0)}=\mp V{\sum_{Z,N}}^{\prime}g\int_{E_0}^{E_s} \!\!dE~\om(A,E) 
\times\nonumber\\
&&\int\!\frac{d^3p}{(2\pi)^3}\, \ln \left( 1\mp
e^{-\beta(p^2/2Am +E-B_{Z,N} -\mu_{Z,N})} \right)\,,
\eea
where the prime on $\sum$ denotes exclusion of the light nuclei $(A \leq 8)$ from 
the sum. The lower limit $E_0$ is determined by the beginning of the excited states
and the applicability of the level density formula at low energy. The upper limit 
$E_s$ is the smallest of separation energies of any particle within the nucleus,
beyond which lies the continuum. We take $E_0=2~$ MeV and $E_s=8~$ MeV.   

We next consider the scattering piece, $\ln {\cz}_{scat}$, which is formally given 
by \cite{Dashen1},
\be
\ln {\cz}_{scat} =\sum \int dE e^{-\beta (E-\mu)}\frac{1}{2\pi i} 
Tr \left({\cal A}S^{-1}(\ep)\frac{\partial}{\partial E}S(\ep)\right)_c
\ee
where the sum is over all scattering channels, each having its chemical potential
$\mu$ and formed by taking any number of particles from any of the stable species 
(proton, neutron and nuclei in their ground and excited states) and the trace is 
over all plane wave states for each of these channels. $S$ is the usual scattering 
operator and ${\cal A}$, the  
boson symmetrization and fermion antisymmetrization operator. The subscript $c$ 
denotes only the connected parts of the expression in parenthesis. To restate 
Eq.(11) explicitly for nuclear matter, we have to characterize the scattering 
channels.

Consider the set of channels, in which the constituting particles have a {\it total}
number of $Z_t$ protons and $N_t$ neutrons with mass number $A_t =Z_t +N_t$. Let 
$\sg$ denote all other labels 
required to fix a channel within this set. Clearly the total mass and the chemical 
potential is independent of $\sg$, depending only on $Z_t$ and $N_t$. If $\vec{P}$ 
is the total momentum of all the particles in a channel, its non-relativistic 
energy is given by 
\be
E_{Z_t,N_t,\sg}=\frac{P^2}{2A_t m} -B_{Z_t,N_t,\sg} +\ep\,,
\ee
where $\ep$ is the kinetic energy in the c.m. frame. Here $B_{Z_t,N_t,\sg}$ is the
sum of individual binding energies of all the particles in the channel. With the 
channels so characterized, one can integrate over $\vec{P}$ to get 
\bea
\ln {\cz}_{scat}&&\!\!\!\!\!\!\!\! = V\!\sum_{Z_t,N_t}\frac{e^{\beta \mu_{Z_t,N_t}}}
{\lm^3(A_tm)}\sum_{\sg} e^{\beta B_{Z_t,N_t,\sg}} \times \nonumber \\
&&\!\!\!\!\!\!\int_0^{\infty}\!\!\!d\ep~e^{-\beta\ep} \frac{1}{2\pi i} 
Tr_{Z_t,N_t,\sg}
\left({\cal A}S^{-1}(\ep)\frac{\partial}{\partial\ep}S(\ep)\right)_c \, ,
\eea
the trace being now restricted to the channel $(Z_t, N_t, \sg)$.

Eq.(13) brings out the nature of convergence of the virial series. Here the integral
depends mildly on temperature. Then the factor $e^{\beta B_{Z_t,N_t,\sg}}$ makes 
channels with large binding energies more important, determining the leading 
behaviour of the series. For an estimate of the convergence domain, we take 
$B_{Z_t,N_t,\sg}\simeq b(Z_t+N_t)$ for such channels with $b \sim 8.5$ MeV, the
binding energy per nucleon of moderately large nuclei. We thus see that Eq.(13) 
may be regarded as a series in powers of 
$e^{\beta(\mu_p+b)}$ and $e^{\beta(\mu_n+b)}$, rather than simply of 
$\ze_p$ and $\ze_n$, as appears from the original definition (3). If we ignore the 
excited states of the nuclei, the series converges for 
$|\ze_p|\,,\,|\ze_n| < e^{-\beta b}\,,$ which, as expected, is smaller than that for 
the ideal gas. Below we shall find a more realistic criterion of convergence.  

Having discussed the convergence of the virial series, we now try to calculate 
$\ln {\cz}_{scat}$ from all the scattering channels. It is convenient to divide the 
channels into {\it light} ones, consisting of low mass particles ($A < 8$, say) 
and {\it heavy} ones, containing at least one high mass particle ($A \geq 8$), so 
that we write
\be
\ln {\cz}_{scat} = \ln {\cz}_{light} + \ln {\cz}_{heavy}\,,
\ee
as the sum of contributions from the {\it light} and the {\it heavy} channels. The 
integrals in Eq.(13) over the $S$-matrix elements appear difficult to evaluate at 
first sight. But the presence of the Boltzmann factor simplifies the task by 
limiting the range of integration to rather low energies, if the temperature is 
small enough.

Generally speaking, two-particle channels are expected to dominate over 
multiparticle channels with same $Z_t$ and $N_t$ from the binding energy 
consideration. As in Ref.\cite{Horowitz} we neglect all light channels, except the 
three two-particle channels, namely, $N\!N,~N\al$ and $\al\al$ to get
\be
\ \ln {\cz}_{light}=\ln {\cz}_{N\!N} + \ln {\cz}_{N\al} + \ln {\cz}_{\al\al}
\ee
We shall write below the individual contributions in terms of phase shifts in 
the respective channels.

The restriction of the integrals in Eq.(13) to the low energy region is all the 
more helpful for estimating the scattering contribution from the heavy channels. 
The cross-sections in these channels are known experimentally to be dominated by 
a multitude of narrow resonances near their thresholds. The $S$-matrix elements 
may thus be well approximated by these resonances. We then have the elegant result 
by Dashen {\it et al} \cite{Dashen2,Dashen3}: The corresponding partition function 
becomes that of a number of ideal gases, one for each of these resonances. 
Assuming their level density to be of the same form as for the excited states 
\cite{Feshbach}, we thus see that $\ln {\cz}_{heavy}$ is again given by the
same formula (10) as for $\ln {\cz}_{ex}^{(0)}$, with the $E$-integral now running 
from $E_s$ to $E_r$, the end of the resonance region. We take $E_r \simeq 12$ MeV. 
We also terminate the series over $Z_t$ and $N_t$ at the same values that we choose
to do for $Z$ and $N$ in Eqs.(8) and (10). Then the two contributions can be 
combined together by extending the variable $E$ from $E_0$ to $E_r$,
\be
\ln {\cz}_{ex}^{(0)}+\ln {\cz}_{heavy}= \mbox{r.h.s. of Eq.(10) with}\, \, 
E_s\rightarrow E_r \,.
\ee

Let us summarize our result at this stage for the grand partition function of 
nuclear matter as, 
\be
\ln {\cz}=\ln {\cz}^{(0)}_{gr} +(\ln {\cz}^{(0)}_{ex} +\ln {\cz}_{heavy}) 
+\ln {\cz}_{light}
\ee
with the terms given by Eqs.(8), (16) and (15) respectively. It is 
observed that the first two terms represent simply the ideal gas terms. 

We now write explicitly the contributions from the three light, two-body channels 
retained in Eq.(15) for $\ln {\cz}_{light}$. These may be recovered from 
Ref.\cite{Horowitz} (see \cite{Comment2}, however), but the master formula (13)
gives them immediately. 
As we are considering elastic two-body scattering, the trace in Eq.(13) 
becomes a sum over the derivative of phase-shifts of the appropriate partial waves.
It gives formulae of the same form as derived by Beth and Uhlenbeck \cite{Beth} for 
the second virial coefficient for particles without spin and isospin. The results 
for each of these channels can be expressed in terms of an integral of the form,
\be
\Delta_{AB}(\beta)=\frac{1}{\pi\,T}\int_0^\infty \!\! d\ep~ e^{-\beta \ep}~
\de_{AB}(\ep),
\ee
where $A$ and $B$ are the scattering particles and $\de_{AB}$ is the sum over  
phase-shifts of the relevant partial waves. We shall also indicate the isospin $I$ 
for the $N\!N$ channel by a superscript on $\Delta$ and $\de_{AB}$. 

For the $N\!N$ channel,
\bea
\ln {\cz}_{N\!N}=&&\frac{V}{\lm^3(2m)} \{ (\ze_p^2+\ze_n^2)\De^{I=1}_{N\!N}\nonumber\\
&&+\ze_p\ze_n ( -3 +\De^{I=1}_{N\!N} + \De^{I=0}_{N\!N})\}
\eea
with  
\be
\de^I_{N\!N}(\ep)=\sum_{S,L,J} (2J+1)\, \de_{{2S+1}_{\textstyle L_{J}}}(\ep)\,.
\ee
Here the contributing partial waves are determined by $I$ through the requirement 
of antisymmetry on the total wave function of the $N\!N$ system. Thus 
\bea
\de^{I=1}_{N\!N}&=&\de_{1_{ \textstyle S_{0}}}+ \de_{3_{ \textstyle P_{0}}}+
3\de_{3_{ \textstyle P_{1}}}+5\de_{3_{ \textstyle P_{2}}}+
5 \de_{1_{ \textstyle D_{2}}}+\cdots\nonumber\\
\de^{I=0}_{N\!N}&=&3\de_{3_{ \textstyle S_{1}}}+ 3\de_{1_{ \textstyle P_{1}}} 
+3\de_{3_{ \textstyle D_{1}}}+5 \de_{3_{ \textstyle D_{2}}}
+7\de_{3_{ \textstyle D_{3}}}+\cdots\nonumber\\
\eea
The term with $-3$ in Eq.(19) arises from partial integration in Eq.(13), 
as the phase-shift for the $^3S_1$ wave (containing the deuteron bound state) 
at threshold is not $0$ but $\pi$, by the Levinson theorem \cite{Goldberger}. 

For the $N\al$ channel \cite{Comment2},
\be
\ln {\cz}_{N\al}=\frac{V}{\lm^3(5m)}\,(\ze_p+\ze_n)\ze_p^2\,\ze_n^2 
\,e^{\beta B_\al}\De_{N\al}\,,
\ee
where $B_\al$ is the binding energy of the alpha-particle and
\bea
\de_{N\al} (\ep)&=&\sum_{L,J} (2J+1)\,\de_{L_{J}} (\ep)\nonumber\\
&=& 2\de_{S_{1/2}}+ 2\de_{P_{1/2}}+ 4\de_{P_{3/2}}+ 4\de_{D_{3/2}}+ 6\de_{D_{5/2}}
+\cdots\nonumber\\
\eea

Lastly we consider the $\al\al$ channel with both spin and isospin equal to zero, 
where 
\be
\ln {\cz}_{\al\al}=\frac{V}{\lm^3(8m)}\,\ze_p^4\,\ze_n^4\, 
e^{2\beta B_\al}\De_{\al\al}\,,
\ee
with
\be
\de_{\al\al} (\ep) =\sum_L (2L+1)\,\de_L (\ep) = \de_S + 5\de_D +9\de_G +\cdots
\ee
The integrals $\De_{AB}(\beta)$ have also been evaluated in Ref.\cite{Horowitz} with
the available phase shifts in the low energy region for all the scattering channels. 

Our result (17) for the grand partition function is not in the form of the virial
series (13). However, we may expand each of the ideal gas terms in it 
again in a virial series following Eq.(4) to get the first two terms as  
\bea
&&\!\!\!\!\!\!\!\!\ln {\cz}_{gr}^{(0)} = V\sum_{Z,N}\frac{g}{\lm^3(Am)}\left(\zeta_{Z,N} \pm
\frac{\zeta^2_{Z,N}}{2^{5/2}} +\cdots\right)\,,\\
&&\!\!\!\!\!\!\!\!\ln {\cz}_{ex}^{(0)}+\ln {\cz}_{heavy} \nonumber\\
&&\!\!\!\!\!\!\!\!= V{\sum_{Z,N}}^{\prime} \frac{g}{\lm^3(Am)}\left(f_1\,\zeta_{Z,N} \pm
f_2\,\frac{\zeta^2_{Z,N}}{2^{5/2}} +\cdots\right)\,,
\eea
in powers of 'effective fugacities', $\zeta_{Z,N}=e^{\beta(\mu_{Z,N}+B_{Z,N})}$,
similar to those already appearing in Eq.(13).
The $A$-dependent constants, $f_n (A),\,n=1,2,\cdots$
\be
f_n(A)=\int_{E_0}^{E_r}\!\! dE~ \om (A,E)~ e^{-n\beta E}\,, 
\ee
are generally large for small $n$, if the temperature is not too low, but 
decreases steadily with increasing $n$, so that they make the series (27) converge
faster. The ideal gas pieces may now be readily calculated, given the binding 
energies of nuclei. We take these binding energies from Ref.\cite{Myers}, where 
they are available for some 9000 nuclei in their ground states up to $Z=135$.

\begin{figure}
\centerline{\psfig{figure=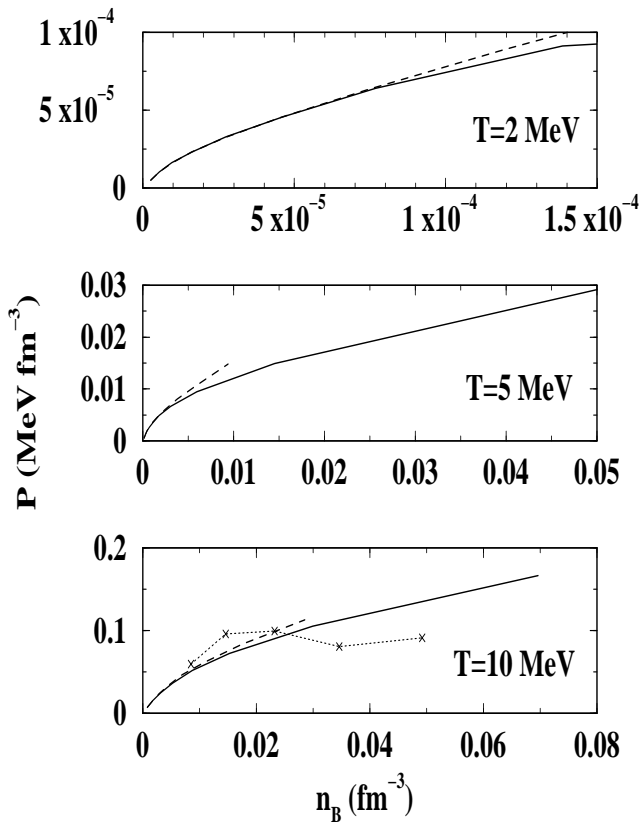,height=11cm,width=8.5cm}}
\caption{Plot of pressure of symmetric nuclear matter against density at $T=2, 5$
and $10$ MeV. The solid curves represent our calculation while the dashed ones are
from the virial calculation of Ref.\cite{Horowitz}, both curves drawn for the same 
range of fugacity at each temperature. The dotted curve with crosses (data 
points) show the microscopic calculation of Ref.\cite{Pan} for $T=10$ MeV.}
\end{figure}

The pressure $P$ and the nucleon densities $n_{p,n}$ are obtained from the
familiar formulae,
\be
P=T \frac{\ln {\cz}}{V}\,, ~~~n_i =\ze_i\left ( \frac{\partial}{\partial \ze_i}
\frac{\ln {\cz}}{V}\right )_{V,T}, ~~i=p,n
\ee
and the total baryon density $n_B$ from $n_B=n_p +n_n$, at different temperatures 
and fugacities.
 
We now consider symmetric nuclear matter, $\ze_p=\ze_n=\ze$ \cite{Comment3}.
At each temperature the numerical evaluation of density and pressure shows a 
steep rise in both the quantities, as we increase the fugacity beyond a certain 
value. We may rely on our evaluation for fugacities below the onset of such rise. 
It turns out that the allowed range of fugacity so determined is also given by half 
the radius of convergence of the virial series obtained earlier in the absence of 
excited states of nuclei, that is, $\ze < (1/2) e^{-\beta b}$.
In Fig.1 we plot pressure against density for $T=2, 5$ and $10$ MeV, where we draw 
the curves for this range of $\ze$ for the first two temperatures, while for the 
third we do so for a smaller range to remain well below the 
saturation density. It is found that for fugacities (densities) considered here, 
nuclei heavier than $A=15$ do not play any significant role.  

For comparison we show in Fig.1
the results of Horowitz and Schwenk \cite{Horowitz}, who calculate $\cz$ 
including only one heavy species, namely the alpha particle, besides the original 
proton and neutron and the low mass scattering channels formed by them, 
denoted here by ${\cz}_{light}$. It is observed that our results 
for pressure, besides being somewhat lower at higher densities, extend over a 
wider region in density than those in Ref.\cite{Horowitz}. The difference is 
to be attributed to nuclei heavier than the alpha particle. Also shown are the 
results of the microscopic calculation of Ref.\cite{Pan} up to a density, beyond 
which the pressure turns out to be unphysical up to a significant region, the 
compressibility there becoming negative.

Our numerical evaluation shows that the contribution from light
scattering channels is about an order of magnitude smaller than that from the
ideal gas terms. We are thus led, to a good approximation, to a {\it modified}
statistical equilibrium model for nuclear matter at moderate temperature, where we 
include the contributions from not only the nuclei and their excited states as in 
the conventional model, but also the heavy scattering channels,
which reduce again to ideal gas terms, once the $S$-matrices in these channels 
are approximated by resonances present in the low energy region.

It remains to calculate other interesting quantities, like the relative abundance 
of nuclei and the energy density and also consider asymmetric systems. Further, 
the results may be extended to higher temperatures by including pions in the initial 
set of species. It is, of course, known that 
chiral symmetry restricts the pions themselves to contribute significantly to the 
pressure \cite{Leutwyler}. But the contribution from their scattering with the 
nucleons and the $\alpha$-particles may not be negligible.

To conclude, we have started from the complete virial series for the grand partition 
function of nuclear matter, expressed entirely in terms of observables. 
Besides the binding 
energies of the nuclei and the density of their excited states, the observables 
include the $S$-matrix elements for scattering of any number of stable nuclear 
particles. While the scattering amplitudes in a few low mass, two-body channels 
are estimated from the experimental scattering data, those in heavy mass 
channels are assumed to be dominated by low energy resonances.     
We thus derive in a simple way the equation of state of nuclear matter at moderate 
temperature, that includes all significant contributions and, at the same time, is 
free from any serious theoretical uncertainty. Further, the results are valid 
up to a higher density than found previously.

\section*{Acknowledgments}
We thank D. Bandyopadhyay and D.N. Basu for discussions.  
S.M. and J.N.D. acknowledge support of DST, Government of India and
S.K.S. of CSIR, Government of India.

\end{document}